\documentstyle[12pt]{article}

\renewcommand{\baselinestretch}{1.85}

\newtheorem{theorem}{Theorem}            
\newcommand{\sds}{\mbox{$\subset \!\!\!\!\!\!+$}}

\begin{document}

\renewcommand{\baselinestretch}{1.1}

\title{\Large \bf Riemannian Space-times 
            of G\"odel Type in Five Dimensions \\} 
\author{
M.J. Rebou\c{c}as\thanks{
{\sc internet: reboucas@cat.cbpf.br}  } \ \ 
and  \  
A.F.F Teixeira\thanks{
{\sc internet: teixeira@novell.cat.cbpf.br}  }  \\ 
\\  
$^{\ast}$ $^{\dagger}~$Centro Brasileiro de Pesquisas F\'\i sicas\\
             \ \ Departamento de Relatividade e Part\'\i culas \\
                 Rua Dr.\ Xavier Sigaud 150 \\
                 22290-180 Rio de Janeiro -- RJ, Brazil \\ \\
       }
        
\date{\today} 

\maketitle

\begin{abstract}
    
The five-dimensional (5D) Riemannian G\"odel-type 
manifolds are examined in light of the equivalence problem 
techniques, as formulated by Cartan.
The necessary and sufficient conditions for local homogeneity 
of these 5D manifolds are derived. 
The local equivalence of these homogeneous Riemannian manifolds 
is studied. It is found that they are characterized by two essential 
parameters $m^2$ and $\omega\,$: identical pairs $(m^2, \, \omega)$ 
correspond to locally equivalent 5D manifolds. An irreducible
set of isometrically nonequivalent 5D locally homogeneous Riemannian 
G\"odel-type metrics are exhibited. A classification of these 
manifolds based on the essential parameters is presented, 
and the Killing vector fields as well as the corresponding Lie algebra 
of each class are determined. It is shown that apart from the 
$(m^2= 4\, \omega^2, \,\omega\not=0)$ and $(m^2\not=0, \, \omega = 0)$ 
classes the homogeneous Riemannian G\"odel-type manifolds admit a 
seven-parameter maximal group of isometry ($G_7$). The special
class $(m^2= 4\,\omega^2, \, \omega\not=0)$ and the degenerated
G\"odel-type class $(\,m^2\not=0,\, \omega=0)$ are shown to have a
$G_9$ as maximal group of motion. The breakdown of causality in
these classes of homogeneous G\"odel-type manifolds are also 
examined.
\end{abstract}

{\raggedright
 \section{Introduction} }      \label{intro}
\setcounter{equation}{0}

Kaluza-Klein-type theories in five and more dimensions has been
of notable interest in several contexts. 
In the framework of gauge theories they have been used in the 
quest for unification of the fundamental interactions in physics.
The idea that the various interactions in nature might
be unified by enlarging the dimensionality of the space-time,
has a long and honourable history that goes back to the works 
of Nordstr\"om, Kaluza and Klein%
~\cite{Nordstrom1914}~--~\cite{Klein1926}.
Its earlier adherents were mainly interested in extending
general relativity, but a late increased interest has been
apparent in the particle physics community, especially among 
those investigating supersymmetry.

The possibility that space-time has more than four dimensions
has also received much attention regarding its cosmological aspects.
Investigation has been focused on attempts to understand, for example,
why the universe presently appears to have only four space-time 
dimensions, and whether it is a higher dimensional dynamically 
evolving manifold ---~the space-time expands while the extra
dimensions contract or remain constant. The first cosmological
model in which the extra dimension contracts as a result of the
cosmological evolution was proposed by Chodos and Detweiler%
~\cite{ChodosDetweiler1980}.  Since then, a great deal of work
has emerged along this line of research, particularly as regards
exact solutions of Einstein's equations, entropy production
during the contracting process~\cite{AlvarezGavela83},
resolution of the horizon and flatness problems~\cite{Guth81}, 
and the like (see~\cite{BurakovskyHorwitz1995} and references
therein). 

{}From a purely technical viewpoint higher-dimensional 
Kaluza-Klein-type theories have also been used as a way of finding 
new exact solutions of Einstein's equations in four dimensions%
~\cite{IbanezVerdaguer1986}~--~\cite{VladimirovKokarev1997}.

More recently Wesson~\cite{Wesson83}~--~\cite{Wesson85} has given a 
new impetus to the study of (4+1)-gravity by investigating 
a five-dimensional extension of general relativity with a variable 
rest mass (see also~\cite{GronSoleng87} and~\cite{Wesson90}). 
In this theory we have space-time-mass (STM) Riemannian manifolds, 
and the fifth dimension is a convenient mathematical way of 
geometrizing the rest mass and  of allowing one to study the 
possibility that it may be variable. 
The four-dimensional (4D) general relativity theory is recovered when
the rate of change of the rest mass is zero.
Ever since the foundations of this Kaluza-Klein-type STM theory 
were laid, there have been investigations on its potentialities 
and physical consequences, particularly as concerns its consistency 
with Mach's principle~\cite{GuangwenMa90}, causality conditions and
inflation~\cite{GronSoleng87,Gron88}, confrontation between theory and 
observation~\cite{Wesson86,Coley90}, and the solutions of its field 
equations~\cite{Wesson90}.

In 1949 G\"odel found a solution of Einstein's field equations
with cosmological constant for incoherent matter with rotation%
~\cite{Godel49}. 
Owing to its striking properties, the cosmological solution
presented by G\"odel has a well-recognized importance and has 
to a large extent motivated the investigations on rotating 
cosmological space-times within the framework of general relativity. 
Particularly, the search for rotating G\"odel-type space-times 
has received a good deal of attention in recent years, and 
the literature on these geometries is fairly large today 
see~\cite{Som68}~--~\cite{Krasinski97} and references therein).

However, the general problem of space-time local homogeneity 
(ST homogeneity, hereafter~\cite{KoikeTamimotoHosoya94})
of four-dimensional Riemannian manifolds 
endowed with a G\"{o}del-type metric was considered for 
the first time only in 1980 by Raychaudhuri and Thakurta%
~\cite{RaychaudhuriThakurta80}. They found the necessary
conditions for ST homogeneity of these manifolds. 
Three years later, Rebou\c{c}as and Tiomno~\cite{ReboucasTiomno83}
proved that Raychaudhuri-Thakurta necessary conditions are also
sufficient for ST homogeneity of G\"odel-type manifolds.
However, in both articles~\cite{RaychaudhuriThakurta80,%
ReboucasTiomno83} the study of ST homogeneity is limited in that 
only time-independent Killing vector fields were considered%
~\cite{TeixeiraReboucasAman85}. 
The Raychaudhuri-Thakurta-Rebou\c{c}as-Tiomno conditions were finally 
proved to be the necessary and sufficient conditions for a 
G\"{o}del-type  manifold to be ST homogeneous without assuming 
any such simplifying hypothesis in~\cite{ReboucasAman87}, where the 
powerful equivalence problem techniques for Riemannian space-times, 
as formulated in terms of spinors by Karlhede~\cite{Karlhede80} 
and embodied in a suite of computer algebra programs 
called {\sc classi}~\cite{Aman87} written in 
{\sc sheep}~\cite{MacCallumSkea94}, were used.

In the light of the equivalence problem techniques, as formulated by
Cartan~\cite{Cartan51} and using {\sc classi}, we extend these
investigations by examining a class of five-dimensional 
(5D) Riemannian G\"odel-type manifolds.
The necessary and sufficient conditions for local homogeneity of these
5D manifolds are derived. The local equivalence of these 
homogeneous space-time-mass Riemannian manifolds 
is discussed; they are found to be characterized by two essential
parameters $m^2$ and $\omega$: identical pairs $(m^2, \,\omega)$ 
correspond to locally equivalent STM manifolds. An irreducible
set of isometrically nonequivalent 5D homogeneous 
Riemannian G\"odel-type metrics are exhibited. 
A classification of these manifolds based on the essential parameters 
is presented, and the Killing vector fields as well as the corresponding 
Lie algebra of each class are determined. We show that apart from 
the $(m^2= 4\, \omega^2, \,\omega\not=0)$ and 
$(m^2\not=0, \, \omega = 0)$ classes the 5D locally homogeneous 
Riemannian G\"odel-type manifolds have a seven-parameter maximal 
group of isometry ($G_7$). The special class 
$(m^2= 4\,\omega^2, \, \omega\not=0)$ and the degenerated
G\"odel-type class $(\,m^2\not=0,\, \omega=0)$ are shown to admit a
$G_9$ as maximal group of motion. The breakdown of causality in
these classes of homogeneous G\"odel-type manifolds is
also examined and shown to have the same basic features of the 
corresponding 4D counterparts.

To close this introduction, we should like to emphasize that although 
the STM Wesson's theory of gravitation is often referred to, the 
results of the following sections hold for any five-dimensional
Riemannian G\"odel-type manifold regardless of the underlying
5D Kaluza-Klein-type theory%
\cite{Wesson1992A}~--~\cite{BillyardWesson1997}
one may be concerned with.

\vspace{4mm}
{\raggedright
\section{Homogeneity, Irreducible Set and Causality} } 
\label{homoconds}
\setcounter{equation}{0}

The arbitrariness in the choice of coordinates in the geometric
theories of gravitation gives rise to the problem of deciding
whether or not two Riemannian manifolds whose metrics $g$ and 
$\tilde{g}$ are given explicitly in terms of coordinates, viz.,
\begin{equation}
ds^2         =        g_{\mu \nu} \,dx^\mu \, dx^\nu \qquad 
\mbox{and} \qquad 
d\tilde{s}^2 = \tilde{g}_{\mu \nu}\,d\tilde{x}^\mu\,d\tilde{x}^\nu\,,
\end{equation}
are locally the same. This is the so-called equivalence problem
(see Cartan~\cite{Cartan51} for the local equivalence of  $n$-dimensional 
Riemannian manifolds, Karlhede~\cite{Karlhede80} and
MacCallum~\cite{MacCallumSkea94,MacCallum83,MacCallum91}
for the special case $n=4$ of general relativity).

\begin{sloppypar}
The Cartan solution~\cite{Cartan51} to the equivalence problem 
for Riemannian manifolds can be reworded as follows. Two 
$n$-dimensional Lorentzian Riemannian manifolds $M$ and 
$\widetilde{M}$ are locally equivalent if there exist coordinate 
and generalized $n$-dimensional Lorentz transformations such that 
the following {\em algebraic} equations relating the frame components 
of the curvature tensor and their covariant derivatives: 
\end{sloppypar}
\parbox{14cm}{\begin{eqnarray*} 
R^{A}_{\ BCD} & = &  \widetilde{R}^{A}_{\ BCD}\,, \nonumber \\
R^{A}_{\ BCD;M_1} & = & \widetilde{R}^{A}_{\ BCD;M_1}\,, \nonumber  \\
R^{A}_{\ BCD;M_1 M_2} &=&\widetilde{R}^{A}_{\ BCD;M_1 M_2}\,,\nonumber  \\
                  & \vdots &   \nonumber \\
R^{A}_{\ BCD;M_1\ldots M_{p+1}} & = & \widetilde{R}^{A}_{\ BCD;M_1
                                         \ldots M_{p+1}} \nonumber \\  
            \end{eqnarray*}}  \hfill
\parbox{1cm}{\begin{eqnarray}   \label{eqvcond}  \end{eqnarray}}
are compatible as equations in  $\left( x^{\mu}, \xi^{A} \right)$. Here 
and in what follows we use a semicolon to denote covariant derivatives.
Note that $x^{\mu}$ are coordinates on the manifold $M$ while $ \xi^{A}$
parametrize the group of allowed frame transformations 
[$n$-dimensional generalized Lorentz group usually denoted%
~\cite{HawkingEllis73} by $O(n-1, 1)\,$]. Reciprocally, equations 
(\ref{eqvcond}) imply local equivalence between the $n$-dimensional 
manifolds $M$ and $\widetilde{M}$.

In practice, a fixed frame is chosen to perform the calculations so 
that only coordinates appear in the components of the curvature tensor,
i.e. there is no explicit dependence on the parameters $\xi^{A}$ of the 
generalized Lorentz group.

Another important practical point to be considered, once one wishes 
to test the local equivalence of two Riemannian manifolds, is that 
before attempting to solve eqs.\ (\ref{eqvcond}) one can extract and 
compare partial pieces of information at each step of differentiation
as, for example, the number $\{t_{0},t_1, \dots ,t_{p}\}$ of 
functionally independent functions of the coordinates $x^\mu$
contained in the corresponding set
\begin{equation}   \label{CartanScl}
I_{p} = \{ R^{A}_{\ BCD} \,, \,R^{A}_{\ BCD;M_{1}} \,, \,
R^{A}_{\ BCD;M_1 M_2}\,,\,\ldots,\,R^{A}_{\ BCD;M_1 M_2\ldots M_{p}}\}\,,  
\end{equation} 
and the isotropy subgroup $\{H_{0}, H_1, \ldots ,H_{p}\}$ of the 
symmetry group $G_r$ under which the set corresponding $I_p$ is 
invariant. They must be the same for each step $q= 0, 1, \cdots ,p$ 
if the  manifolds are locally equivalent.

In practice it is also important to note that in calculating the 
curvature and its covariant derivatives, in a chosen frame, one 
can stop as soon as one reaches a step at which the $p^{th}$ 
derivatives (say) are algebraically expressible in terms of the 
previous ones, and the residual isotropy group  (residual frame 
freedom) at that step is the same isotropy group of the previous 
step, i.e.  $H_p = H_{(p-1)}$. In this case further differentiation
will not yield any new piece of information.
Actually, if $H_p = H_{(p-1)}$ and, in a given frame, the $p^{th}$ 
derivative is expressible in terms of its predecessors, for any 
$q > p$ the $q^{th}$ derivatives can all be expressed in terms 
of the $0^{th}$, $1^{st}$, $\cdots$, $(p-1)^{th}$ derivatives%
~\cite{Cartan51,Thomas34,MacCallumSkea94}.  
As in the worst case we have only one functionally independent function 
of the coordinates $x^\mu$ at each step of the differentiation process, 
and the generalized Lorentz group has $n (n-1)/2$ independent parameters, 
it follows that for five-dimensional Riemannian manifolds $p+1 \leq 15$.

Since there are $t_p$  essential coordinates, in 5D 
clearly  $5-t_p$ are ignorable, so the isotropy group will 
have dimension  $s = \mbox{dim}\,( H_p )$, and the group of 
isometries of the metric will have dimension $r$ given by%
(see Cartan~\cite{Cartan51})
\begin{equation}
r = s + 5 - t_p \,, \label{gdim}
\end{equation}
acting on an orbit with dimension
\begin{equation}
d = r - s = 5 - t_p \,.  \label{ddim}
\end{equation}

The line element of the five-dimensional Riemannian 
G\"odel-type manifolds $M_5$ we are concerned with in this work 
is given by 
\begin{equation} \label{ds2} 
ds^{2} = dt^2 + 2\,H(x)\, dt\,dy - dx^2 - G(x)\,dy^2 - dz^2 - du^2\,,
\end{equation}
where $H(x)$ and $G(x)$ are arbitrary real functions of $x$, and the 
five STM coordinates clearly are  $t, x, y, z, u$.  As a matter of
fact, to ensure the local Lorentzian character of (\ref{ds2})
one has to require that $H^2(x) + G\,(x) > 0$.
At an arbitrary 
point of $M_5$ one can choose the following set of linearly 
independent one-forms $\theta^A$: 
\begin{equation} \label{lorpen}
\theta^{0} = dt + H(x)\,dy\,, \: \quad
\theta^{1} = dx\,, \: \quad
\theta^{2} = D(x)\,dy\,, \:\quad
\theta^{3} = dz \,, \: \quad
\theta^{4} = du \,,          
\end{equation}
such that the G\"odel-type line element (\ref{ds2}) can be written
as
\begin{equation} \label{ds2f}
ds^2 = \eta_{AB} \: \theta^A \,\, \theta^B = (\theta^0)^2 
- (\theta^1)^2 - (\theta^2)^2 - (\theta^3)^2 - (\theta^4)^2\,, 
\end{equation}
where $D^2(x) = G + H^2$. Here and in what follows capital letters
are pentad indices (or Lorentz frame indices) and run from 0 to 4;
they are raised and lowered with Lorentz matrices
$\eta^{AB} = \eta_{AB} = \mbox{diag} (+1, -1, -1, -1, -1)$,
respectively. 

Using as input the one-forms (\ref{lorpen}) and the Lorentz frame
(\ref{ds2f}), the computer algebra package {\sc classi} gives the 
following nonvanishing Lorentz frame components $R_{ABCD}$ of the 
curvature:
\begin{eqnarray}  
R_{0101} &=& R_{0202}=- \frac{1}{4} \, \left( \frac{H'}{D}\, \right)^2\,, 
                        \label{rie1st} \\  
R_{0112} & =& \frac{1}{2} \, \left( \frac{H'}{D}\, \right)' \,, 
                         \label{rie2nd}\\  
R_{1212} &=& \frac{D''}{D}-\frac{3}{4}\,
            \left( \frac{H'}{D}\,\right)^2 \label{rielast}\,,
\end{eqnarray}
where the prime denotes derivative with respect to $x$.

For STM homogeneity from eq.\ (\ref{ddim}) one must have 
$t_q=0$ for $q=0, 1, \cdots\, p$, that is, the number of 
functionally independent functions of $x^\mu$ in
the set $I_p$ must be zero. Therefore, from 
eqs.\ (\ref{rie1st})~--~(\ref{rielast})
we conclude that for STM homogeneity it is necessary that
\begin{eqnarray}
\frac{H'}{D} &=&\mbox{const} \equiv -\,2\,\omega \label{stmcond1} \,, \\
\frac{D''}{D}&=&\mbox{const} \equiv m^2 \,. \label{stmcond2}
\end{eqnarray}

We shall now show that the above necessary conditions are also
sufficient for STM local homogeneity. Indeed, under these conditions
the nonvanishing frame components of the curvature reduce to
\begin{eqnarray}  
R_{0101} &=& R_{0202}=- \omega^2 \label{rieh1st} \,, \\  
R_{1212} &=& m^2 - 3\,\omega^2 \label{riehlast} \,.
\end{eqnarray}
Following Cartan's method for the local equivalence,
we next calculate the first covariant derivative of
the Riemann tensor. Now one obtains the following non-null
covariant derivatives of the curvature:
\begin{equation} \label{drieh}
R_{0112;1} = R_{0212;2}= \omega\, (m^2  - 4\,\omega^2) \,. 
\end{equation}
As the first covariant derivative of the curvature is 
algebraically expressible in terms of the Riemann tensor and
for a given pair $(m^2, \omega)$ the isotropy group $H_1$ is 
the same as $H_0$ (see next paragraph), no new covariant 
derivative of the curvature should be calculated. Clearly one 
also has $t_0 = t_1 =0$.

As far as the dimension of the residual isotropy group is 
concerned we distinguish two different classes of locally homogeneous
5D G\"odel-type Riemannian manifolds, according to the relevant
parameters $m^2$ and $\omega$, namely~\cite{footnote1}
\begin{enumerate}
\item[(i)]
$\, m^2 \not= 4\,\omega^2\,$ and $\,\omega \not=0$ for 
which $\,\mbox{dim}\,(H_0) = \mbox{dim}\, (H_1)= 2\,$;
\item[(ii)]
$m^2 = 4\, \omega^2$ with $\,\omega \not=0$, and the degenerated 
G\"odel-type manifolds $m^2 \not= 0$ and $\omega =0$. Here one has
$\,\mbox{dim}\,(H_0) = \mbox{dim}\, (H_1)= 4\,$.
\end{enumerate}

Thus, from eqs.\ (\ref{gdim}) and (\ref{ddim}) 
one finds that the locally homogeneous 5D Riemannian G\"odel-type 
manifolds admit a (local) $G_r$, with either $r =7$ or $r=9$, acting 
on an orbit of dimension 5, that is on the whole STM manifold.

The above results can be collected together in the following theorems:

\vspace{2mm} 
\begin{theorem} \label{TheoHom}
The necessary and sufficient conditions for a five-dimensional 
Riemannian G\"odel-type manifold to be locally homogeneous 
are those given by  equations (\ref{stmcond1})~--~(\ref{stmcond2}).
\end{theorem}
\begin{theorem} \label{GroupTheo}
The five-dimensional locally homogeneous Riemannian G\"o\-del-type 
manifolds admit group of isometry $G_r$ with 
\begin{enumerate}
\item[(i)]
$r=7\:$  if $\:\, m^2 \not= 4\,\omega^2\,$ and $\,\omega \not=0$;
\item[(ii)]
$r=9\:$ if $\:\,m^2 = 4\, \omega^2$ with $\omega \not=0$, or when  
$m^2 \not= 0$ and $\,\omega =0$.
\end{enumerate}
\end{theorem}
\begin{theorem} \label{EquivTheo}
The five-dimensional homogeneous Riemannian G\"o\-del-type 
manifolds are locally characterized by two independent 
parameters $m^2$ and $\omega$: identical pairs 
($m^2, \omega$) specify locally equivalent manifolds.
\end{theorem}

We remark that the particular case $m^2=\omega=0$ has not been 
included in our study inasmach as, from (\ref{rieh1st}) and 
(\ref{riehlast}), it is clearly the 5D flat manifold.

We shall now be concerned with the irreducible set of 
isometrically nonequivalent homogeneous G\"odel-type metrics. 
To this end, we distinguish four classes of metrics according to:

{\bf Class I} : $\,m^2 > 0, \: \omega \not=0 $. 
For this case, the general solution of (\ref{stmcond1}) and 
(\ref{stmcond2}) can be written as
\begin{equation}
D(x) = a_0\, e^{mx} + a_1 \, e^{-mx}
\qquad  \mbox{and} \qquad  
H(x) = -\, \frac{2\,\omega\,}{m}\: (a_0\, e^{mx} - a_1\, e^{-mx} ) + a_2 \,,
\end{equation} 
where $a_0$, $a_1$ and $a_2$ are arbitrary constants.
According to the above theorem (\ref{EquivTheo}), the constants 
$a_0$, $a_1$ and $a_2$ are not essential. In other words 
they can be eliminated by coordinate transformations.
Indeed, if one  performs the coordinate transformation $u' =u$ 
and successively the transformations (3.10), (3.15) and 
(3.16) of~\cite{ReboucasTiomno83} one finds that the line element 
for this class of homogeneous G\"odel-type manifolds is brought 
into the form 
\begin{equation} \label{ds2c}
ds^{2}=[\,dt+H(r)\, d\phi\,]^{2} -D^{2}(r)\, d\phi^{2} -dr^{2} 
                 -dz^{2} - du^2
\end{equation}
in cylindrical coordinates $(r, \phi, z)$, where
\begin{equation} \label{DHe}
H(r) =\frac{2\,\omega}{m^{2}}\: [1 - \cosh\,(mr)] 
\qquad \mbox{and} \qquad
D(r) = \frac{1}{m} \sinh\,(mr) \,.
\end{equation}

{\bf Class II} : $\,m^2 = 0, \: \omega \not=0 $. 
For this case, the general solution 
of (\ref{stmcond1}) and (\ref{stmcond2}) is
\begin{equation}
D(x) = b_0\,x - b_1 
\qquad \mbox{and} \qquad 
H(x) = -\omega\,x\, \left( b_0 x - 2 b_1 \right) + b_2 \,,
\end{equation}
where $b_0, b_1$ and $b_2$ are arbitrary constants. By trivial
coordinate transformations the line element for this class 
can be brought to the form~(\ref{ds2c}) but now with     
\begin{equation} \label{DHsr}
H(r) = - \,\omega\, r^{2}  \: \qquad \mbox{and} \: \qquad D(r) = r \,,
\end{equation}
where only the essential parameter $\omega$ appears 
($m^2=0$, for this class).

{\bf Class III} : $\,m^{2} \equiv - \mu^{2} < 0, \: \omega \not=0 $. 
Similarly for this class, the integration of the conditions for homogeneity
(\ref{stmcond1}) and (\ref{stmcond2}) leads to
\begin{equation} 
D(x) = c_0\, \sin\,(\mu x )+ c_1 \, \cos\,(\mu x )
\quad \mbox{and} \quad  
H(x) = \,\frac{2\,\omega\,}{\mu} \,
\left[\,c_0\,\cos\,(\mu x) - c_1\, \sin\,(\mu x)\, \right] + c_2 \,.
\end{equation} 
Here again the non-essential constants $c_0, c_1$ and $c_2$
can be eliminated by coordinate transformations so that the 
line element for this class reduces to~(\ref{ds2c}) with
\begin{equation}   \label{DHc}      
H(r) = \frac{2\,\omega}{\mu^{2}} \:[\cos\,(\mu r) - 1 ]
\qquad \mbox{and} \qquad  
D(r) = \frac{1}{\mu} \sin\,(\mu r)\,.
\end{equation}

{\bf Class IV} : $\,m^{2} \not= 0, \: \omega = 0 $. We refer to 
the manifolds of this class as degenerated G\"odel-type manifolds,
since the cross term in the line element, related to the
rotation $\omega$ in G\"odel model, vanishes. 
By a trivial coordinate transformation one can make $H = 0$ 
with $D(r)$ given, respectively, by (\ref{DHe}) 
or (\ref{DHc}) depending on whether $m^2>0$ or
$m^{2} \equiv - \mu^{2} < 0$. 

In three out of the above four classes of  homogeneous 
G\"odel-type manifolds there are closed timelike curves. 
Indeed, the G\"odel's analysis for 4D manifolds can be 
extended to the 5D manifolds in a straightfoward way to 
prove this. To this end, we write the line element~(\ref{ds2c}) 
as
\begin{equation} \label{ds2cx} 
ds^2=dt^2 +2\,H(r)\, dt\,d\phi -dr^2 -G(r)\,d\phi^2 -dz^2 -du^2\,,
\end{equation}
where $G(r)= D^2 - H^2$ and $(r, \theta, \phi)$ are cylindrical 
coordinates.

The existence of closed timelike curves of the G\"odel-type depends
on the behavior of $G(r)$. Indeed, if  $G(r) < 0$ for a certain
range of $r$ ($r_1 < r < r_2$, say), G\"odel's circles%
~\cite{CalvaoRevoucasTeixeiraSivaJr88} $u,t,z,r =const$ are 
closed timelike curves. 

It is, therefore, easy to show that for the above class II 
there is a critical radius $r_c$, defined by 
$|\omega|\, r_c=1$, such that for $r_c < r < \infty$ the
circles $u, t, z = const$ and $r = const > r_c$ are 
closed timelike curves. So there is breakdown of causality 
for this class.
 
For the class III noncausal G\"odel circles occur depending on the
root $r_n$ of the equation
\begin{equation}
\sin^2 \left[ \frac{\mu r_n}{2} \right] = 
\left[ 1+ \frac{4\,\omega^2}{\mu^2} \right]^{-1} \leq 1 \,.
\end{equation} 
Thus, for this class there are alternatively causal and noncausal
regions (circular rings) in the surfaces $u, t, z = const$. 

For the class I if $m^2 < 4\, \omega^2$ the noncausal G\"odel 
circles occur for $ r > r_1$, where
\begin{equation}
\sinh^2 \left[ \frac{m r_1}{2} \right] =
\left[ 1+ \frac{4\,\omega^2}{m^2} \right]^{-1} \geq 1 \,.
\end{equation}
However, for $m^2 \geq 4\,\omega^2$ then $G(r) > 0$ and there is 
no breakdown of causaly of G\"odel-type. As a matter of fact, 
following Calv\~ao's reasoning~\cite{CalvaoRevoucasTeixeiraSivaJr88} 
one can show that: 
({\em a\/}) although causal, the $m^2= 4\,\omega^2$ manifolds are not 
stably causal~\cite{HawkingEllis73}, that is, an arbitrary small
perturbation of this metric gives rise to causality violation;
({\em b\/}) the family $m^2>4\,\omega^2$ is stably causal, which
means that the causal properties of this family are invariant
under arbitrary small perturbation of its metrics.
We remark that Calv\~ao's approach holds only if the 5D homogeneous
Riemannian G\"odel-type manifolds are endowed with the topology
of ${\cal R}^5$.
 
Finally, for the degenerated G\"odel-type geometries
(class IV), as $G(r)= D^2 >0$ there is no closed timelike 
curves of G\"odel type.

\vspace{4mm}
{\raggedright
\section{Isometries} } \label{isomet}  
\setcounter{equation}{0}

In this section we shall be concerned with the isometries
of the homogeneous 5D Riemannian G\"odel-type manifolds,
whose line element (\ref{ds2c}) can be brought into the
Lorentzian form (\ref{ds2f}) but now with the one-forms 
$\theta^A$ given by
\begin{equation} \label{lorpen1}
\theta^{0} = dt + H(r)\,d\phi\,, \: \quad
\theta^{1} = dr\,, \: \quad
\theta^{2} = D(r)\,d\phi\,, \:\quad
\theta^{3} = dz \,, \: \quad
\theta^{4} = du \,,
\end{equation}
where the functions $H(r)$ and $D(r)$ are given by
(\ref{DHe}), (\ref{DHsr}) or (\ref{DHc}) depending on
the sign of $m^2$.

Denoting the coordinate components of a generic Killing vector field 
$K$ by $K^{\mu} \equiv (Q, R, S, Z, U)$,
where $Q, R, S, Z$ and $U$ are functions of all coordinates 
$t,r,\phi,z$, $u$, then the fifteen Killing equations 
\begin{equation} \label{killeqs}
         K_{(A;B)} \equiv K_{A;B} + K_{B;A} = 0 
\end{equation}
can be written in the Lorentz frame (\ref{ds2f})~--~(\ref{lorpen1}) as
\begin{eqnarray}  
&T_t  =  0 \,, \qquad    T_z - Z_t  =  0 \label{um} \,, \\
&R_r  =  0 \,, \qquad   Z_r + R_z  =  0  \label{dois} \,, \\
&Z_z  =  0  \,,   \label{tres} \\
&D\,(T_r - R_t) - H_r P  =  0  \label{quatro} \,,          \\
&D P_z + Z_{\phi} - H Z_t  =  0 \label{cinco} \,,        \\
&T_{\phi} + H_r R - D P_t  =  0 \label{seis}  \,,        \\
&R_{\phi} - H R_t - D_r P + D P_r  =  0  \label{sete} \,, \\
&P_{\phi} - H P_t + D_r R  =  0  \label{oito}        \,, \\ 
&T_{u} - U_t  =  0  \label{nove}   \,, \\
&R_{u} + U_r  =  0  \label{dez}    \,, \\
&D P_{u} - H U_t + U_{\phi}  =  0  \label{onze}   \,, \\
&U_z + Z_{u}  =  0   \label{doze}  \,, \\
&U_{u}  =  0   \label{treze}  \,,
\end{eqnarray}
where the subscripts denote partial derivatives, and where we 
have made
\begin{equation}  \label{pencor}
T \equiv H\,S + Q \qquad \quad \mbox{and} \quad \qquad P \equiv D\,S
\end{equation}
to make easier the comparison and the use of the results obtained 
in~\cite{TeixeiraReboucasAman85}. To this end we note that in the
present form the equations (\ref{um})~--~(\ref{oito}) are formally
identical to the Killing equations (4) to (11) of%
~\cite{TeixeiraReboucasAman85}. However, in the above equations
(\ref{um})~--~(\ref{oito}) 
the functions $T, R, P, Z$ depend additionally on the fifth 
coordinate $u$. Taking into account this similitude, 
the integration of the Killing equations (\ref{um})~--~(\ref{treze}) 
can be obtained in two steps as follows. 
First, by analogy with (4) to (11) of~\cite{TeixeiraReboucasAman85}
one integrates (\ref{um})~--~(\ref{oito}), but at this step instead of 
the integration constants one has integration functions of the fifth
coordinate $u$. 
Second, one uses the remaining eqs.\ (\ref{nove})~--~(\ref{treze})
to achieve explicit forms for these integration functions and
to obtain the last component $U$ of $K$.

In what follows we shall use the above two-steps procedure 
to integrate the Killing equations (\ref{um})~--~(\ref{treze}) 
for the class I of the section 2 in details. 
However, for the sake of brevity, for the remaining classes
we shall only present the Killing vector fields and the 
corresponding Lie algebras without going into details 
of calculation, which can be similarly obtained, or simply verified 
by using the computer algebra program {\sc killnf},
written in {\sc classi} by {\AA}man~\cite{Aman87}, with further 
extensions by MacCallum~\cite{MacCallumSkea94}. As a matter of fact, 
even the Killing equations (\ref{um})~--~(\ref{treze}) have been 
achieved by using {\sc killnf}.

{\bf Class I} : $\,m^2 > 0, \: \omega \not=0 $. In the integration of
the Killing equation for this general class one is led to distinguish
two different classes of solutions depending on whether 
$m^2 \not= 4 \,\omega^2$ or $m^2 =4 \,\omega^2$. We shall refer to these
subclasses as classes Ia and Ib, respectively. 

{\bf Class Ia} : $\,m^2 > 0, \: m^2 \not= 4\,\omega^2 $. According to 
the above two-steps procedure, from (\ref{pencor}) and according to
eqs.\ (36)~--~(39) of~\cite{TeixeiraReboucasAman85} one finds
\begin{eqnarray}
Q &=& \alpha + 2\,\omega \left( \gamma\,D_{r} + \delta\, D \sin \phi
                        + \epsilon\, D \cos \phi \right) \nonumber \\
  & & +\, \frac{H}{D}\, \left( \gamma\,m^2 D + \delta\, D_r \sin\phi
         + \epsilon\, D_r \cos\phi \right) \,,  \label{Qeq} \\
R &=& \delta\, \cos\phi - \epsilon\, \sin\phi \,, \label{Req} \\
S &=& - \,\gamma\, m^2 - \frac{D_r}{D}\, \left( \delta\,\sin\phi 
         + \epsilon \, \cos\phi \right) \,, \label{Seq} \\
Z &=& \beta \,, \label{Ueq} \\ 
U &=& U(t,r,\phi, z,u) \label{Veq}\,,
\end{eqnarray}
where $\alpha, \beta, \gamma, \delta$ and $\epsilon$ are integration
functions of the fifth coordinate $u$.

{}From (\ref{doze}) and ({\ref{treze}) one finds
\begin{eqnarray}   
\beta &=& \kappa_7 \, u + \kappa_2 \,, \label{Beteq} \\
U     &=& -\,\kappa_7 \, z + \zeta (t, r, \phi) \,, \label{Veq1}
\end{eqnarray}
where $\kappa_2$ and $\kappa_7$ are arbitrary constants and
$\zeta (t,r, \phi)$ is an arbitrary function.

Equation (\ref{dez}) can now be used to obtain the integration 
functions $\delta$ and $\epsilon$ and to simplify the expression
of $U$. In doing so one finds
\begin{eqnarray} 
&\delta    =  \delta_0 \, u + \kappa_4 \,, \quad \qquad  
\epsilon  =  \epsilon_0 \, u + \kappa_5 \,, \label{Epseq1} \\
&U     = - \,\kappa_7 \, z - \delta_0\, r \cos\phi 
        + \epsilon_0\, r \sin\phi + \nu (t,\phi) \,, \label{Veq2}
\end{eqnarray}
where $\delta_0, \epsilon_0, \kappa_4, \kappa_5$ are constants
and $\nu\,(t,\phi) $ is an arbitrary function.

Finally, equations (\ref{nove}) and (\ref{dez}) give the remaining
integration functions $\alpha$, $\gamma$ and reduce the functions
$\delta$, $\epsilon$ and $U$ to
\begin{eqnarray} 
\gamma   & = & const \equiv m^{-1}\, \kappa_3  \,,  \qquad \:    
\alpha    =  const \equiv \kappa_1  \,,   \label{GamAlp}    \\
\delta   & = & \kappa_4   \,,  \quad \:  
\epsilon  =  \kappa_5   \,,  \quad  \:
U     = - \,\kappa_7 \, z + \kappa_6 \,, \label{Veq3}
\end{eqnarray}
where $\kappa_6$ is a new arbitrary constant.

Inserting (\ref{Beteq}) and (\ref{GamAlp})~--~(\ref{Veq3}) into
(\ref{Qeq})~--~(\ref{Veq}) and using (\ref{DHe}) one obtains
\begin{eqnarray}
Q &=& \kappa_1 + \frac{2\,\omega}{m}\,\kappa_3\,- \,\frac{H}{D}\, 
\left(\kappa_4\,\sin \phi +\kappa_5\, \cos \phi \right) 
              \label{Qeq1} \,,  \\
R &=& \kappa_4 \, \cos\phi - \kappa_5\, \sin\phi \,, \label{Req1} \\
S &=& - \,\kappa_3\, m -  \frac{D_r}{D}\left(\kappa_4\,\sin\phi 
         + \kappa_5 \,\cos\phi \right) \,, \label{Seq1} \\
Z &=& \kappa_7\, u + \kappa_2 \,, \label{Ueq1} \\ 
U &=& -\,\kappa_7\, z + \kappa_6 \label{Veq4} \,.
\end{eqnarray}

Thus, in the coordinate basis in which as (\ref{ds2c}) is given,
a set of linearly independent Killing vector fields 
$K_N$ ($N$ is an enumerating index) 
can be written as
\begin{eqnarray}
K_1 &=&\partial_t \,,  \quad \: 
K_2 = \partial_z  \,,  \quad  \: 
K_3 = \frac{2\,\omega}{m}\, \,\partial_t - m \,\partial_{\phi} \,, 
           \label{KIa1} \\
K_4 &=& -\,\frac{H}{D}\, \sin\phi\, \,\partial_t +\cos\phi\, \,\partial_r
        -\,\frac{D_r}{D}\, \sin\phi\, \,\partial_{\phi} \,, 
                          \label{KIa2} \\
K_5 &=& -\,\frac{H}{D}\, \cos\phi\, \,\partial_t -\sin\phi\, \,\partial_r
        -\,\frac{D_r}{D}\, \cos\phi\, \,\partial_{\phi} \,, 
           \label{KIa3} \\
K_6 &=& \partial_{u} \,, \qquad \quad 
K_7 = u\,\,\partial_z - z\,\,\partial_{u} \,. \label{KIa4}
\end{eqnarray}

The Lie algebra has the following nonvanishing commutators:
\begin{eqnarray}
&\left[ K_2, K_7 \right]    =  - K_6  \,, \qquad 
\left[ K_3, K_4 \right] = - m\, K_5   \,, \\
& \left[ K_3, K_5 \right] = m\, K_4  \,, \qquad
\left[ K_4, K_5 \right]    =  m\, K_3 \,, \\   
&\left[ K_6, K_7 \right]  =   K_2   \,. 
\end{eqnarray} 
So, the corresponding algebra is 
${\cal L}_{Ia} = t^2 \,\sds \, so\,(2) \oplus \tau \oplus so\,(2,1)$.
The semi-direct sum of sub-algebras $t^2 \,\sds \, so\,(2)$ 
corresponds to the translations $K_2$ and $K_6$, and the  rotation $K_7$.
The symbol $\tau$ is associated to the time translation $K_1$. Finally, 
the infinitesimal generators of sub-algebra $so\,(2,1)$ are 
$K_3,\, K_4\,$ and $K_5$.
 
{\bf Class Ib} : $\,m^2 =4\, \omega^2 \,, \:\, \omega\not=0 $. For this
class, using the results of~\cite{TeixeiraReboucasAman85} and
(\ref{pencor}) one finds
\begin{eqnarray}
Q &=& \alpha + \gamma - \frac{H}{D} \left( \delta \sin \phi
     + \epsilon  \cos \phi \right) 
   - \frac{H}{D}\, \left[\eta \sin(mt + \phi) 
         + \xi \cos(mt + \phi) \right] \,,  \label{QQeq} \\
R &=& \delta\, \cos\phi - \epsilon\, \sin\phi 
   + \xi \sin(mt+\phi) - \eta \cos(mt+\phi) \,, \label{RReq} \\
S &=& - \gamma\, m - \frac{D_r}{D} \left( \delta \sin\phi 
         + \epsilon \cos\phi \right) 
      + \frac{1}{D} \left[ \eta \sin(mt + \phi)
         + \xi \cos(mt + \phi) \right]  \,, \label{SSeq} \\
Z &=& Z(u) \,, \label{UUeq} \\ 
U &=& U(t,r,\phi, z,u) \,,  \label{VVeq}
\end{eqnarray}
where $\alpha, \gamma, \delta$, $\epsilon, \xi $ and $\eta$ are 
(new) integration functions of the fifth coordinate $u$.

Again, from (\ref{doze}) and ({\ref{treze}) one finds
\begin{eqnarray}   
Z &=& k_8 \, u + k_2 \,, \label{Ufun} \\
U     &=& -\,k_8 \, z + \chi\, (t, r, \phi) \,, \label{VVeq1}
\end{eqnarray}
where $k_2$ and $k_8$ are  arbitrary constants and
$\chi (t,r, \phi)$ is an arbitrary function.

Similarly to the class Ia, eqs. (\ref{dez}) can be used to obtain 
a first expression for the integration functions $\xi$, $\eta$ 
$\delta$, $\epsilon$, and to reduce $U$ according to
\begin{eqnarray} 
&\xi = \xi_0 \, u + k_6 \,, \qquad \qquad  
\eta = \eta_0\, u + k_7 \,, \label{Etaeq} \\
&\delta = \delta_0 \, u + k_4 \,, \qquad \qquad 
\epsilon  =  \epsilon_0 \, u + k_5 \,, \label{Epseq2} 
\end{eqnarray}
\begin{equation}
\!\!\!\!U     = - k_8\, z + r \left[ - \delta_0\cos\phi 
     + \epsilon_0 \sin\phi - \xi_0 \sin(mt+\phi) 
     + \eta_0 \cos(mt+\phi) \right]      
  + \:\: \sigma\, (t,\phi) \,,  \label{Veq5}
\end{equation} 
where $\xi_0, \eta_0, \delta_0, \epsilon_0$, $k_4, \cdots ,k_7$ 
are constants, and $\sigma\, (t,\phi)$ is an arbitrary 
function. 

Equations (\ref{nove}) and (\ref{dez}) give the remaining
integration functions $\alpha$, $\gamma$ and reduce the 
functions $\xi, \eta$, $\delta$, $\epsilon$ and $U$ to
\begin{eqnarray} 
\xi & =&  k_6 \,, \label{xiconst} \quad                      
\eta = k_7 \,, \quad                         
\gamma    =  const \equiv k_3  \,, \quad    
\alpha    =  const \equiv k_1  \,, \label{ALPHA} \\
\delta  & = & k_4   \,,    \quad            
\epsilon  =  k_5   \,,     \quad            
U     = - \,k_8 \, z + k_9 \,, \label{Veq6}
\end{eqnarray}
where $k_9$ is a new arbitrary constant.

Inserting (\ref{Ufun}) and (\ref{xiconst})~--~(\ref{Veq6}) into
(\ref{QQeq})~--~(\ref{VVeq}) one obtains
\noindent
\begin{eqnarray}
&&\!\!\!\!\!\!\!\!\!\!\!\!\!\!\!\!\!
Q =k_1+k_3 - \frac{H}{D} \left( k_4 \sin \phi
     + k_5  \cos \phi \right)  
     - \,\frac{H}{D}\, \left[k_7 \sin(mt + \phi) 
         + k_6 \cos(mt + \phi) \right] \,,  \label{QQeq1} \\
&&\!\!\!\!\!\!\!\!\!\!\!\!\!\!\!\!\!
R = k_4\, \cos\phi - k_5\, \sin\phi 
   + k_6 \sin(mt+\phi) - k_7 \cos(mt+\phi) \,, 
         \label{RReq1}   \\
&&\!\!\!\!\!\!\!\!\!\!\!\!\!\!\!\!\!
S = - k_3 \, m - \frac{D_r}{D} \left( k_4 \sin\phi 
         + k_5 \cos\phi \right)  
     +\, \frac{1}{D}\, \left[ k_7 \sin(mt + \phi)
         + k_6 \cos(mt + \phi) \right]  \,, \label{SSeq1}  \\
&&\!\!\!\!\!\!\!\!\!\!\!\!\!\!\!\!\!
Z = k_8\, u + k_2  \,,  \label{UUeq1}  \\ 
&&\!\!\!\!\!\!\!\!\!\!\!\!\!\!\!\!\!
U = -\, k_8\, z + k_9 \,,  \label{VVeq2} 
\end{eqnarray}
which give rise to the Killing vector fields
\begin{eqnarray}
K_1 &=&\partial_t \,,  \quad \: 
K_2 = \partial_z  \,,  \quad  \: 
K_3 = \partial_t - m \,\partial_{\phi} \,, 
           \label{KIb1} \\
K_4 &=& -\,\frac{H}{D}\, \sin\phi\, \,\partial_t +\cos\phi\, \,\partial_r
     -\,\frac{D_r}{D}\, \sin\phi\, \,\partial_{\phi} \,, \label{KIb2} \\
K_5 &=& -\,\frac{H}{D}\, \cos\phi\, \,\partial_t -\sin\phi\, \,\partial_r
        -\,\frac{D_r}{D}\, \cos\phi\, \,\partial_{\phi} \,, \label{KIb3} \\
K_6 &=&-\,\frac{H}{D}\,\cos(mt+\phi)\,\,\partial_t 
          +\sin(mt+\phi)\,\,\partial_r
    +\,\frac{1}{D}\, \cos(mt+\phi)\,\,\partial_{\phi} \,, \label{KIb4} \\
K_7 &=&-\,\frac{H}{D}\,\sin(mt+\phi)\,\,\partial_t 
            -\cos(mt+\phi)\,\,\partial_r
    +\,\frac{1}{D}\, \sin(mt+\phi)\,\,\partial_{\phi} \,, \label{KIb5} \\
K_8 &=& u\,\,\partial_z - z\,\,\partial_{u} \,, \qquad \quad 
K_9 = \partial_{u} \,, \label{KIb6} 
\end{eqnarray}
whose Lie algebra is given by
\begin{eqnarray}
&\left[ K_1, K_6 \right] = -m\, K_7 \,, \qquad
\left[ K_1, K_7 \right] = m\, K_6 \,, \qquad
\left[ K_2, K_8 \right]    =  - K_9  \,,  \\ 
&\left[ K_3, K_4 \right] = - m\, K_5   \,, \qquad  
\left[ K_3, K_5 \right] = m\, K_4  \,, \qquad 
\left[ K_4, K_5 \right]    =  m\, K_3 \,, \\   
&\left[ K_6, K_7 \right]  = m\,  K_1  \,, \qquad 
\left[ K_8, K_9 \right]  = - K_2   \,. 
\end{eqnarray} 
So, the corresponding algebra for this case is 
${\cal L}_{Ib} = t^2 \,\sds \, so\,(2) \oplus so\,(2,1) \oplus so\,(2,1)$.
The semi-direct sum $t^2 \,\sds \, so\,(2)$ 
comprises the translations $K_2$ and $K_9$, and the  rotation $K_8$, while
the two sub-algebras $so\,(2,1)$ are generated by the Killing vector
fields $K_1, K_6, K_7$ and  $K_3, K_4, K_5$.

{\bf Class II} : $\,m^2 = 0, \: \omega \not=0 $. 
Similarly for this class, using~(\ref{DHsr}) and the results of
~\cite{ReboucasTiomno83} the above two-steps procedure
gives rise to the following Killing vector fields:
\begin{eqnarray}
K_1 &=&\partial_t \,,  \quad \: 
K_2 = \partial_z  \,,  \quad  \: 
K_3 = \partial_{\phi} \,, \label{KII1} \\
K_4 &=& -\,\omega\,r\, \sin\phi\, \,\partial_t -\cos\phi\, \,\partial_r
    +\,\frac{1}{r}\, \sin\phi\, \,\partial_{\phi} \,, \label{KII2} \\
K_5 &=& -\,\omega\,r\, \cos\phi\, \,\partial_t +\sin\phi\, \,\partial_r
        +\,\frac{1}{r}\, \cos\phi\, \,\partial_{\phi} \,, \label{KII3} \\
K_6 &=& u\,\,\partial_z - z\,\,\partial_{u} \,, \qquad \quad
K_7 = \partial_{u} \,. \label{KII4} 
\end{eqnarray}
The Lie algebra has the following nonvanishing commutators:
\begin{eqnarray}
&\left[ K_2, K_6 \right]    =  - K_7  \,, \qquad 
\left[ K_3, K_4 \right] =  K_5   \,, \\
&\left[ K_3, K_5 \right] = - K_4  \,, \qquad 
\left[ K_4, K_5 \right]    =  2\, \omega\,  K_1 \,, \\   
&\left[ K_6, K_7 \right]  = - K_2   \,. 
\end{eqnarray} 
Therefore, the corresponding algebra for this case is 
${\cal L}_{II} = t^2 \,\sds \, so\,(2) \oplus {\cal L}_4$.
The semi-direct sum of sub-algebras corresponds here to
the translations $K_2$ and $K_7$, and the rotation $K_6$.
The sub-algebra ${\cal L}_4$ is generated by $K_1, K_3,
K_4$ and $K_5$. This algebra ${\cal L}_4$ is soluble and 
does not contain abelian 3D sub-algebras; it is classified
as type $III$ with $q=0$ by Petrov~\cite{Petrov69}.  

{\bf Class III} : $\,m^{2} \equiv - \mu^{2} < 0, \: \omega \not=0 $. 
Again for this class using~(\ref{DHc}) and~\cite{TeixeiraReboucasAman85}, 
the above outlined two-step procedure furnishes
\begin{eqnarray}
K_1 &=&\partial_t \,,  \quad \: 
K_2 = \partial_z  \,,  \quad  \: 
K_3 = \frac{2\,\omega}{\mu} \, \partial_t 
             + \mu\, \partial_{\phi} \,, \label{KIII1} \\
K_4 &=& -\,\frac{H}{D}\, \sin\phi\, \,\partial_t 
            +\cos\phi\, \,\partial_r
   -\,\frac{D_r}{D}\, \sin\phi\, \,\partial_{\phi} \,, \label{KIII2} \\
K_5 &=& -\,\frac{H}{D}\, \cos\phi\, \,\partial_t 
            -\sin\phi\, \,\partial_r
   -\,\frac{D_r}{D}\, \cos\phi\, \,\partial_{\phi} \,, \label{KIII3} \\
K_6 &=& u\,\,\partial_z - z\,\,\partial_{u} \,, \qquad \quad 
K_7 = \partial_{u} \,. \label{KIII4}
\end{eqnarray}
The Lie algebra has the following nonvanishing commutators:
\begin{eqnarray}
&\left[ K_2, K_6 \right]    =  - K_7  \,, \qquad 
\left[ K_3, K_4 \right] =  \mu \, K_5   \,, \\
&\left[ K_3, K_5 \right] = -\mu \, K_4  \,, \qquad 
\left[ K_4, K_5 \right]    =  \mu \, K_3 \,, \\   
&\left[ K_6, K_7 \right]  = -  K_2   \,. 
\end{eqnarray} 
Thus, the corresponding algebra for this case is 
${\cal L}_{III} = t^2 \,\sds \, so\,(2) \oplus \tau \oplus so\,(3)$.
The semi-direct sum of sub-algebras corresponds again to
the translations $K_2$ and $K_7$, and the rotation $K_6$.
Here $\tau$ is associated to the Killing vector field $K_1$,
whereas to the sub-algebra $so\,(3)$ correspond $K_3$,
$K_4$ and $K_5$. 

{\bf Class IV} : $\,m^{2} \not= 0, \: \omega = 0 $. 
This class corresponds to the so-called degenerated 
G\"odel-type manifolds. By a similar procedure one
obtains for this class the following Killing vector
fields:
\begin{eqnarray}
K_1 &=&\partial_t \,,  \quad \: 
K_2 = \partial_z  \,,  \quad  \: 
K_3 = z\,\partial_t + t\, \partial_z \,, \label{KIV1} \\
K_4 &=& \cos\phi\, \,\partial_r
   -\,\frac{D_r}{D}\, \sin\phi\, \,\partial_{\phi} \,, \label{KIV2} \\
K_5 &=& -\sin\phi\, \,\partial_r
  -\,\frac{D_r}{D}\, \cos\phi\, \,\partial_{\phi} \,, \qquad \:
K_6 = \partial_{\phi} \,, \label{KIV3} \\
K_7 & = &u\,\,\partial_z - z\,\,\partial_{u} \,, \quad 
K_8 = u\,\,\partial_t + t \,\,\partial_{u} \,, \quad 
K_9 = \partial_{u} \,, \label{KIV4} 
\end{eqnarray}
where $D(r) = (1/m)\, \sinh mr$ for $m^2>0\,$,  or 
$\,D(r) = (1/\mu)\, \sin \mu r$ for $m^2 \equiv - \mu^2 < 0 $.
The Lie algebra has the following nonvanishing commutators:
\begin{eqnarray}
\left[ K_1, K_3 \right]  &=&  K_2  \,,  \quad 
\left[ K_1, K_8 \right]   =  K_9  \,,   \quad 
\left[ K_2, K_3 \right]   =  K_1  \,,  \\
\left[ K_2, K_7 \right] & = & - K_9  \,,   \quad 
\left[ K_3, K_7 \right]   =  - K_8  \,,    \quad 
\left[ K_3, K_8 \right]   =  - K_7  \,,    \\
\left[ K_4, K_5 \right]  & = & -m^2\, K_6 \,,   \quad
\left[ K_4, K_6 \right]   =  - K_5 \,,         \quad
\left[ K_5, K_6 \right]   =   K_4 \,,          \\   
\left[ K_7, K_8 \right] & = & - K_3   \,,   \quad
\left[ K_7, K_9 \right]   =  - K_2   \,,    \quad
\left[ K_8, K_9 \right]   =  - K_1   \,,
\end{eqnarray} 
where one should substitute $-m^2$ by $\mu^2$ if
$m^2 < 0 $.
So, when $m^2 >0$ the corresponding algebra is 
${\cal L}_{IV} = t^3 \,\sds \, so\,(2,1) \oplus so\,(2,1)$,
where to $t^3$ correspond $K_1, K_2$ and $K_9$.
The infinitesimal generators of $so\,(2,1)$ in the semi-direct 
sum are $K_3, K_7$ and $K_8$, while to the other $so\,(2,1)$
correspond $K_4, K_5$ and $K_6$.
When $m^2 <0$ the algebra associated to $K_4, K_5$ and $K_6$
is $so\,(3)$, instead.

It is worth noting that none of the above Lie algebras
is semi-simple, but some of their sub-algebras are.
Besides, most of the simple sub-algebras are non-compact.
The 3D sub-algebra $so\,(3)$ present in the cases
${\cal L}_{III}$ and ${\cal L}_{IV}$ with $m^2 <0$ are
compact, though.

We have therefore succeeded in finding the maximal group of
motions of all classes of 5D homogeneous Riemannian
G\"odel-type manifolds we have studied in the previous
section.

Equations (\ref{KIa1})~--~(\ref{KIa4}), (\ref{KIb1})~--%
~(\ref{KIb6}), (\ref{KII1})~--~(\ref{KII4}),
(\ref{KIII1})~--~(\ref{KIII4}), and (\ref{KIV1})~--~(\ref{KIV4})
make explicit that the 5D locally homogeneous Riemannian G\"odel-type
manifolds admit maximal group of isometry $G_r$ with: ({\em a\/})
$r=7\:$  if $\: m^2 \not = 4\,\omega^2\,$ and $\,\omega \not=0$;
or ({\em b\/})
$r=9\:$ if $\:\,m^2 = 4\, \omega^2$ with $\omega \not=0$, or when  
$m^2 \not= 0$ and $\,\omega =0$, in agreement with 
theorem~\ref{GroupTheo} of the previous section. Actually the
integration of the Killing equations constitutes a different
way of deriving that theorem. Furthermore, these equations 
also show that the isotropy subgroup $H$ of $G_r$ 
is such that $\,\mbox{dim}\,(H) = 2\,$ for 
$\,m^2 \not= 4\,\omega^2\,$ and $\,\omega\not=0$, while for
$m^2 = 4\, \omega^2$ with  $\,\omega \not=0$ as well as 
for the degenerated G\"odel-type ($m^2 \not= 0$ and $\omega =0$) 
we have $\,\mbox{dim}\,(H) = 4\,$, also in agreement with the 
previous section.   

To conclude, we should like to emphasize again that the results 
of this work hold for any five-dimensional locally homogeneous
Riemannian G\"odel-type manifold, regardless of the underlying
5D Kaluza-Klein-type theory of gravitation one may be
concerned with. This gives a measure of the generality 
of our results in the context of these theories, in which 
the 5D Riemannian manifolds are the underlying arena for 
the formulation of the physical laws.

\vspace{8mm}
{\raggedright
\section*{Acknowledgement} } \label{acknowl}  
\setcounter{equation}{0}

The authors gratefully acknowledge financial assistance from CNPq.


\begin{thebibliography}{99}
\bibitem{footnote1}  In the next section we integrate the Killing 
equations for the 5D Riemannian G\"odel-type manifolds and explicitly 
show that the isotropy group depends on these special relations between
the essential parameters $m^2$ and $\omega$.
\end{thebibliography}
\end{document}